# Quantum size effects in layered VX$_2$ (X=S, Se, Te) materials: Manifestation of metal to semimetal or semiconductor transition

A. H. M. Abdul Wasey, Soubhik Chakrabarty and G. P. Das[*]

Department of Materials Science, Indian Association for the Cultivation of Science, Jadavpur, Kolkata-700032, India

* Corresponding author's email: msgpd@iacs.res.in

**ABSTRACT:** Most of the 2D transition metal dichalcogenides (TMDC) are nonmagnetic in pristine form. However, 2D pristine VX$_2$ (X=S, Se, Te) materials are found to be ferromagnetic. Using spin polarized density functional theory (DFT) calculations, we have studied the electronic, magnetic and surface properties of this class of materials in both trigonal prismatic *2H*- and octahedral *1T*-phase. Our calculations reveal that they exhibit materially different properties in those two polymorphs. Most importantly, detailed investigation of electronic structure explored the quantum size effect in *2H*-phase of these materials thereby leading to metal to semimetal (*2H*-VS$_2$) or semiconductor (*2H*-VSe$_2$, *2H*-VTe$_2$) transition when downsizing from bilayer to corresponding monolayer.

## I. INTRODUCTION

In this era of miniaturization of electronic devices, achieving multi-functionality in low dimensional device materials becomes one of the main objectives of both scientists as well as industries. Low dimensional functional materials are important prerequisites for portable and flexible nano-electronic devices. The seminal discovery of single atomic layer of graphite i.e. graphene [1-2] and two dimensional dichalcogenides (TMDC) like MoS$_2$ [3-4] are proven to be one of the most important milestones in this circumstances. These 2D materials not only provide light weight compact devices because of their low dimensionality, they also offer intriguing new properties and strikingly different physical phenomena at reduced dimension which are absent in their bulk counterpart. Quantum confinement of electron is responsible for these novelties in materials properties at atomic scale which is important from fundamental as well as technological point of view. Despite the remarkable progress in nanoelectronics, realization of spintronic devices with 2D TMDC is a challenging task because most of the pristine 2D TMDCs are nonmagnetic. Other members of this TMDC family namely VS$_2$ and VSe$_2$, structurally similar to MoS$_2$, are well known since 70's because of their charge density wave (CDW) and versatility in electronic structures; however these materials lost attentions for almost three decades presumably for difficulties in synthesis processes [5-11]. Recently VS$_2$ has been experimentally synthesized successfully [12-15] offering an avenue for realization of TMDC based spintronic devices because subsequent theoretical study indicated pristine VS$_2$ to be ferromagnetic in nature [16]. They also proposed that its magnetic properties can be tuned by applying in-plane strain [16]. Along with its novel magnetic properties [16-18], very recent theoretical studies indicate the possibility of its diverse applications as a 'catalyst for electrochemical hydrogen production' [19], '2D anode material for Li ion batteries' [20] etc. In fact, metallic ultra-thin $VS_2$ nanosheet has actually been demonstrated to be a potential 2D conductor for in-plane supercapacitors [13]. Also experimentalists predicted $VS_2$ as a potential material for its moisture responsiveness [21] and efficient field emission properties [15]. However, despite its intriguing electronic properties, sufficient theoretical and experimental results are not available in the literature on this class of $VX_2$ (X=S, Se, Te) materials.

In this paper, we report our first principles density functional theory (DFT) based investigation on the electronic and magnetic properties of $VX_2$ materials which can exist both in trigonal prismatic *2H*- and octahedral *1T*-phase [22]. We found that these exhibit materially different electronic, surface and magnetic properties in their two polymorphs. Studying their electronic structure in monolayer and bilayer, we have theoretically demonstrated quantum size effect in *2H*-phase of these materials resulting into metal to semimetal or semiconductor transition while going from $VX_2$-bilayer to monolayer. This study may have implications on tuning its properties, considering the size of the materials as a controllable parameter.

## II. THEORETICAL METHODOLOGY

$VX_2$ (X=S, Se, Te) can exist in two phases *viz.* trigonal prismatic *2H*-phase and octahedral *1T*-phase where V is octahedrally coordinated with X ions (see FIG. 1). In both the cases V is sandwiched between two X layers and these composite X-V-X trilayers are stacked together one above another *via* weak van der Waals force. For DFT simulation of monolayers and bilayers of $VX_2$, we have incorporated a vacuum slab of ~20 Å in the perpendicular direction to the surface to ensure that there is no interaction with its image in that direction as shown in FIG. 1(a) and (d). Thus Bloch-periodicity is retained in two dimensions and is broken in the perpendicular direction because of incorporation of large vacuum slab. Point to be mentioned here is that, in the entire discussion, by $VX_2$-monolayer we actually mean an X-V-X trilayer of $VX_2$.

The geometry relaxation, total energy calculations and the electronic structure calculations have been performed using spin polarized DFT based Vienna *Ab Inito* Simulation Package (VASP) code [23]. The exchange-correlation functional has been approximated by generalized gradient approximation (GGA) of Perdew-Burke-Ernzerhof (PBE) [24]. The projector augmented wave (PAW) [25] method was employed for describing the electron-ion interactions for the elemental constituents. The plane wave basis cut off was 500 eV for all the calculations performed in this work. Ionic relaxations are performed using conjugant gradient (CG) minimization method [26] to minimize the Hellman-Feynman forces among the constituent atoms with the tolerance of 0.001 eV/Å. For all the total energy calculations reported in this paper, self-consistency has been achieved with a 0.001 meV ($10^{-6}$ eV) convergence. Brillion Zone (BZ) has been sampled using Monkhorst-Pack method [27]. For bulk and slab geometry, 45×45×21 and 45×45×1 k-mesh have respectively been used for electronic structure calculations. As the inter-planar interaction is weak van der Waals (vdW) type, we have taken into account vdW corrections by using DFT-D2 method of Grimme as implemented in VASP code [28].

Main focus of our present work is to study the monolayers and bilayers of $VX_2$ materials which are purely 2D systems. We have therefore estimated two quantities namely surface energy and work function that has important role on governing the properties of a solid surface. Surface energy, which dictates the stability of a 2D material, has been estimated by the following relations, $\sigma = ½ (E_{slab} − (N_{slab}/N_{bulk})E_{bulk})$, where $E_{slab}$ and $E_{bulk}$ are respectively the total energies of slab i.e. monolayer or bilayer and bulk whereas $N_{slab}$ and $N_{bulk}$ represent the corresponding number of atoms in their respective geometry. On the other hand work function can be defined as the minimum amount of energy required to remove an electron from the highest occupied orbital of any solid surface.

## III. RESULTS AND DISCUSSIONS

### A. Structural Properties

First of all we have established the ground state geometry of the bulk $VX_2$ by minimizing the inter-atomic forces by relaxing all the atoms. We have considered both *2H*- and *1T*-phase of $VX_2$ in our investigations (see FIG. A1 in APPENDIX). The lattice parameters *a* and *c* of *1T*-$VS_2$ are estimated to be 3.17 and 5.91 Å which are in reasonably good agreement with available experimental values 3.22 and 5.76 Å respectively [8]. In case of *2H*-phase the estimated numbers come out to be 3.17 and 12.13 Å. The point to be mentioned here is that in case of *2H*-phase the *c* value gets almost doubled because of *AB*-stacking as compared to its *AA*-stacking in *1T*-phase. It is also worth mentioning that in our bulk and multilayer calculations we have considered *AA*-stacking for *1T*-phase and *AB*-stacking for *2H*-phase. This choice indeed has been justified by our total energy considerations. As the inter-layer interaction is weak van der Waals type, we have duly taken care of dispersion corrections. Starting lattice parameters for monolayers and bilayers systems were set to the corresponding bulk values. Estimated lattice parameters for $VX_2$-monolayers and bilayers are shown in TABLE I and TABLE II respectively.

In order to have insight into the natural existence of $VX_2$ in the two phases, we have estimated their relative energetic stability *via* calculating their cohesive energy per $VX_2$ formula unit by the following relation, $E_C = E_{VX2} − (E_V + 2E_X)$, where $E_{VX2}$, $E_V$ and $E_X$ corresponds to the total energies of $VX_2$ system, a single V atom and a single X atom respectively. Cohesive energies of *2H*- and *1T*-$VX_2$-monolayers and bilayers are given in the tables. *2H*-$VS_2$-monolayer is found to be energetically more favorable than its corresponding *1T*-phase marginally by ~26 meV. This trend is followed by $VSe_2$ also albeit with smaller margin of ~8 meV. However, in case of VTe$_2$, *2H*-phase is less favorable than its *1T*-phase by ~52 meV. This overall trend is repeated in case of VX$_2$-bilayers. However, the differences in cohesive energies between *2H*- and *1T*-phase are too small to predict one particular phase precisely to be energetically more superior over another. We have carried out a comparative study of the tendency of VX$_2$ materials to form monolayer or bilayer *via* estimation of surface energy. According to the definition given in the previous section, larger the surface energy value lesser is the possibility to form 2D nanostructure from its bulk counterpart. The estimated surface energy increases down the series i.e. from VS$_2$ to VTe$_2$ monolayers thereby suggesting the fact that VS$_2$-monolayer formation is relatively easier than that of VTe$_2$. Now, as far as the two polymorphs of VX$_2$ are concerned, monolayer formation is slightly more feasible for *2H*-phase than *1T*-phase that is clear from the surface energies of all VX$_2$-monolayers given in TABLE I. In order to have an insight into the size effect we have repeated this investigation on VX$_2$-bilayers and corresponding parameters are given in TABLE II. Interface energies of the bilayers have been estimated by the relation, $E_{int}=E_{bilayer}-2E_{monolayer}$, where $E_{bilayer}$ and $E_{monolayer}$ are the total energies of the corresponding VX$_2$-bilayers and monolayers respectively. Higher the interface energy lesser will be the possibility to reduce its dimension. Thus increasing surface energies are justified by the increase in interface energy down the series. Interface energies are marginally smaller in case of *2H*-phase than that of *1T*-phase for all the VX$_2$-bilayers which is further supported by the slightly larger interlayer distances. From the above investigation it can be argued that *2H*-phase of VX$_2$ materials have more tendencies to form lower dimensional nanostructure than that of its *1T*-phase.

**B. Work Function**

In order to investigate the surface properties of this class of materials we have estimated their work functions. We found that the work function values for both *2H*- and *1T*-VX$_2$-monolayers go on decreasing while going down the series. Moreover, work functions of *2H*-VX$_2$-monolayers are significantly higher than that of *1T*-VX$_2$-monolayers. For example estimated work function of *2H*-VS$_2$-monolayer is 5.87 eV whereas for *1T*-phase the value is found to be 5.45 eV. This difference in work function can be qualitatively rationalized by their difference in electronic structures as shown in FIG. 2 (c) and (d). We have repeated this investigation for VX$_2$-bilayers and found over all similar trend like its monolayer counterpart albeit with slightly different values. This study may have implications on photoelectric and field emission properties of these materials.

**C. Magnetic Properties**

Our spin polarized DFT calculations indicate that VX$_2$-monolayers are magnetic in nature. Estimated magnetic moments at V-site are increasing down the series from *1T*-VS$_2$ (0.51 $\mu_B$) to *1T*-VTe$_2$ (0.93 $\mu_B$) (see TABLE I). This trend is retained in case of *2H*-phase of VX$_2$-monolayers. However interestingly, *2H*-VX$_2$-monolayers have much higher magnetic moment than that of *1T*-VX$_2$-monolayers e.g. in case of VS$_2$ estimated V-moment is 0.51 $\mu_B$ in *1T*-phase whereas in *2H*-phase it is 1.01 $\mu_B$.

This difference in magnetism between the two phases has been supported by the spin density iso-surface (keeping same value for both the phases) plots (see FIG. 2 (a) and (b)) for *2H*- and *1T*-VS$_2$-monolayers. Small magnetic moment is induced in S sites also but in opposite polarity as compared to V-moment which is illustrated by blue (up spin) and pink (down spin) surfaces. This is a manifestation of through-bond spin polarization in VS$_2$-monolayers. We have also carried out this investigation on VX$_2$-bilayers and found somewhat similar results albeit with slightly different total magnetic moment per VX$_2$ formula unit. This marginal difference in total magnetic moment is presumably because of weak inter-layer interaction in bilayers which is absent in the case of monolayers.

**D. Electronic Structures and Quantum Size Effects**

To look into the electronic structure of VS$_2$-monolayers in more detail we first analyzed their electronic density of states (DOS) along with projections (PDOS) on V and S sites (see FIG. 2 (c) and (d)). Strong hybridization of V-*3d* with S-*3p* can be seen from the PDOS plots which indeed are responsible for the strong in-plane covalent bonding. Careful observation of the DOS of *2H*-VS$_2$-monolayer reveals the absence of any states right at the Fermi level, indicating its semimetallic nature unlike the corresponding *1T*-phase. In case of *1T*-VS$_2$-monolayer, metallic nature is clearly evident from the presence of delocalized states near Fermi level originating predominantly from the V-*3d* orbitals, as expected. Larger spin splitting of *2H*-VS$_2$-monolayer compared to its *1T*- counterpart is clearly envisaged from the extent of asymmetry between the up and down spin DOS, as shown in the figures. This indeed corroborates with the higher V-moment in *2H*-phase than *1T*-phase, as discussed in *Sec. C*.

The semimetallic nature of *2H*-VS$_2$-monolayer is in sharp contrast to its bulk counterpart which happens to be metallic (see FIG. A2 in APPENDIX). Question that naturally arises is about the effect of dimensionality (2D to 3D) on the electronic structure of *2H*-VS$_2$. What happens when we add say one more layer of *2H*-VS$_2$? When does the transition from semimetallicity to metallicity take place? We have therefore looked into the evolution of the electronic band dispersion when one goes from monolayer to bilayer. The energy band structure plotted along the high symmetry directions of the 2D hexagonal Brillouin zone (FIG. 3 (a)) clearly shows the valence band top (at Γ point) and the conduction band minimum (between Γ and M point) touching the Fermi level, thereby revealing semimetallic nature of *2H*-VS$_2$-monolayer. It is worth mentioning here that the highest occupied level and the lowest unoccupied level (highlighted bands) have opposite spin polarization, which are labeled by blue (up spin) and red (down spin) color (see FIG. 3 (a)). However, our DFT calculation for the bilayer of *2H*-VS$_2$ clearly show crossing of bands between Γ and M point (see FIG. 3 (c)) indicating the metallic nature as also seen from the total DOS shown in FIG. 3 (d).

Now let us look at the isostructural *2H*-VSe$_2$ and *2H*-VTe$_2$-monolayers. Unlike *2H*-VS$_2$-monolayer, these are found to be narrow indirect band gap semiconductors with the respective gap values of ~0.24 eV (see FIG. 4 (a) and (b)) and ~0.18 eV (see FIG. A4 in APPENDIX). This result is in sharp contrast to their *1T*-counterparts which are found to be metallic (see FIG. A5 and A6 in APPENDIX). Here also, the bilayers of *2H*-VSe$_2$ (see FIG. 4 (c) and (d)) and *2H*-VTe$_2$ (see FIG. A4 in APPENDIX) become metallic because of bands crossing the Fermi level. Thus the metal to semimetal/semiconductor transition while going from bilayer to monolayer *2H*-VX$_2$ (X=S, Se. Te) can be attributed to the enhanced quantum confinement of electron in two-dimension. However, in case of the *1T*-phase quantum confinement effect is unable to make such kind of alteration in its electronic structure while one goes to monolayer (see FIG. 2 (d) and FIG. A7 in APPENDIX).

Then, we look into the case when we modify the bilayer by replacing one of the two VX$_2$ monolayers by MoX$_2$ (X=S, Se) monolayer. The heterobilayers *2H*-MoS$_2$-*2H*-VS$_2$ [29] and *2H*-MoSe$_2$-*2H*-VSe$_2$ are constructed from the corresponding lattice matched (see APPENDIX) monolayers. Both *2H*-MoS$_2$ and *2H*-MoSe$_2$-monolayers are direct band gap semiconductors with respectively ~1.88 eV and ~1.55 eV gap values [30-32]. The results of our self-consistent electronic band structure and DOS are shown in the FIG. 5. In panel (a) the highlighted bands corresponding to up and down spin channels are qualitatively similar to that of free standing *2H*-VS$_2$-monolayer (FIG. 3 (a)). However in presence of *2H*-MoS$_2$-monolayer, those bands are now crossing the Fermi level constituting a metallic nature (see FIG. 5 (a) and (b)) similar to that observed in case of *2H*-VS$_2$-bilayer (FIG. 3 (c) and (d)). Similar effects have been observed for *2H*-MoSe$_2$-*2H*-VSe$_2$-heterobilayer albeit with less prominence as shown in FIG. 5 (c) and (d). We therefore have established the effect of quantum confinement on the electronic structure of *2H*-VX$_2$-bilayers as well as of *2H*-MoX$_2$-*2H*-VX$_2$-heterobilayers.

**IV. CONCLUSION**

In summary, we have carried out first principles spin polarized DFT calculations on monolayer, bilayer and heterobilayer of VX$_2$ (X=S, Se, Te) family. Our calculations reveal that, *2H*-phases have stabilities comparable to their *1T*-counterpart as revealed by total energy considerations. Work function values of *2H*-phase are found to be considerably higher than that of *1T*-phase. These two phases are found to have different spin splitting, *2H*-phase having higher magnetic moment than *1T*-phase. These differences have been justified by their electronic structures. From the detailed analysis of electronic structure of *2H*-VX$_2$-monolayers and bilayers, we have established the evidence of quantum size effects. Metallic *2H*-VX$_2$-bilayers become semimetallic (VS$_2$) or semiconducting (VSe$_2$ and VTe$_2$) while transiting to monolayer which can be explained from the significant alteration of the degree of electron confinement. Thus our systematic first principle investigations demonstrate the phase and size dependent properties of this VX$_2$-family that may have implications over using these materials for possible device applications.

**ACKNOWLEDGEMENT**

This work has been carried out partially under the IBIQuS project which provides financial support to AHMAW. SC is financially supported by a CSIR fellowship 09/080(0787)/2011-EMR-I. GPD gratefully acknowledges the financial support received from the Dept. of Atomic Energy, Govt. of India (DAE) for the IBIQuS project. AHMAW and SC thank Dr. Debjani Karmakar, Dr. Ranjit Thapa and Rajib Batabyal for many helpful discussions during this work.

**TABLES**

TABLE I. Some estimated physical parameters are tabulated for $VX_2$-monolayers. In magnetic moment column total magnetic moment is given and within parenthesis moment at V-site is being provided.

| Materials | Energetics (eV) | | Lattice parameter $a$ (Å) | Magnetic moment ($\mu_B$) | Work function (eV) |
|---|---|---|---|---|---|
| | Cohesive energy ($E_C$) | Surface energy ($\sigma$) | | | |
| $1T$-$VS_2$-monolayer | -14.8849 | 0.0745 | 3.17 | 0.48 (0.51) | 5.45 |
| $1T$-$VSe_2$-monolayer | -13.5086 | 0.1056 | 3.32 | 0.59 (0.65) | 4.95 |
| $1T$-$VTe_2$-monolayer | -12.0279 | 0.1616 | 3.58 | 0.82 (0.93) | 4.54 |
| $2H$-$VS_2$-monolayer | -14.9114 | 0.0710 | 3.17 | 1.00 (1.01) | 5.87 |
| $2H$-$VSe_2$-monolayer | -13.5169 | 0.0998 | 3.33 | 1.00 (1.08) | 5.51 |
| $2H$-$VTe_2$-monolayer | -11.9761 | 0.1584 | 3.58 | 1.00 (1.11) | 4.97 |

TABLE II. Some estimated physical parameters are tabulated for $VX_2$-bilayers. In magnetic moment column total magnetic moment is given and within parenthesis moment at V-site is being provided.

| Materials | Energetics (eV) | | | Structural details(Å) | | Magnetic moment ($\mu_B$) | Work function (eV) |
|---|---|---|---|---|---|---|---|
| | Cohesive energy ($E_C$) | Interface energy ($E_{int}$) | Surface energy ($\sigma$) | Lattice parameter ($a$) | Interlayer distance ($d$) | | |
| $1T$-$VS_2$-bilayer | -14.9566 | -0.1433 | 0.0774 | 3.18 | 3.01 | 0.89(0.48) | 5.53 |
| $1T$-$VSe_2$-bilayer | -13.6112 | -0.2051 | 0.1086 | 3.32 | 3.14 | 1.19(0.66) | 4.99 |
| $1T$-$VTe_2$-bilayer | -12.1845 | -0.3132 | 0.1666 | 3.59 | 3.31 | 1.75(0.98) | 4.55 |
| $2H$-$VS_2$-bilayer | -14.9801 | -0.1374 | 0.0732 | 3.17 | 3.10 | 1.83(0.93) | 5.98 |
| $2H$-$VSe_2$-bilayer | -13.6143 | -0.1948 | 0.1022 | 3.32 | 3.20 | 1.91(1.03) | 5.36 |
| $2H$-$VTe_2$-bilayer | -12.1278 | -0.3035 | 0.1651 | 3.58 | 3.36 | 1.92(1.07) | 4.86 |

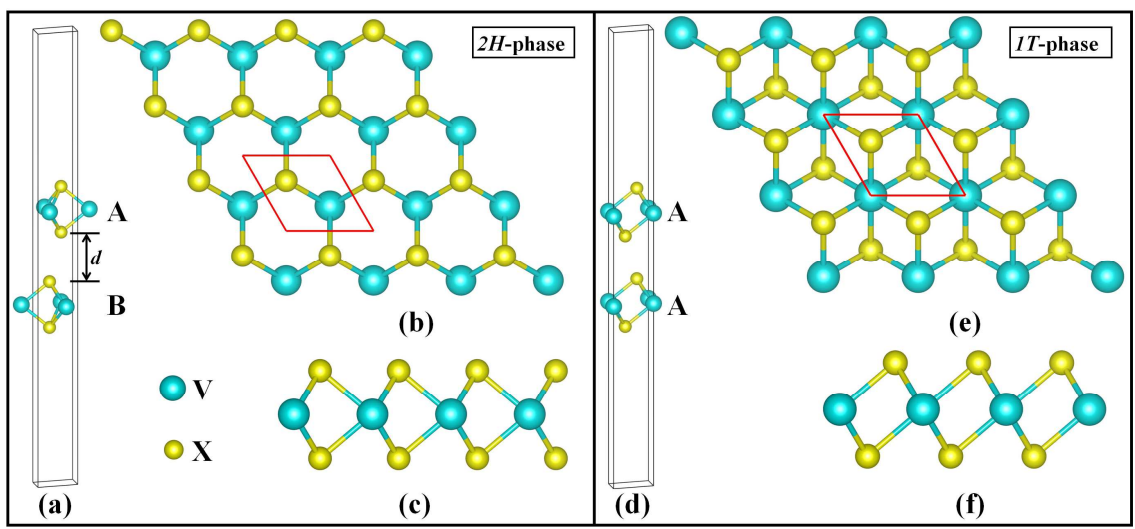

FIG. 1. Two different polymorphs of MX$_2$. AB-type stacking of *2H*-phase of MX$_2$-bilayer is shown in (a). Top view (b) and side view (c) of the MX$_2$-monolayer are shown. Figure (d)-(f) show the respective structures for *1T*-phase of MX$_2$.

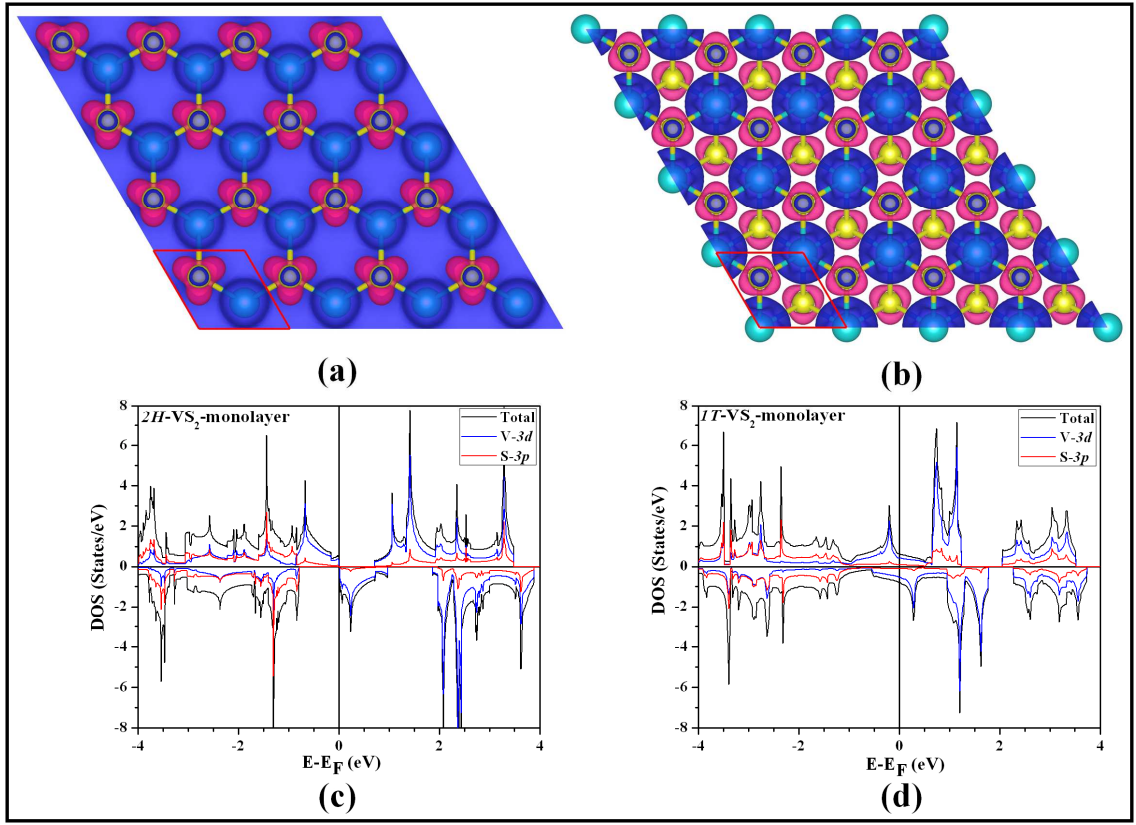

FIG. 2. Showing the spin density iso-surface plots (keeping the same value of 0.002 e/Å$^3$ for both the plots) for (a) *2H*-VS$_2$-monolayer and (b) *1T*-VS$_2$-monolayer. Also total and projected DOS of (c) *2H*-VS$_2$-monolayer and (d) *1T*-VS$_2$-monolayer are shown.

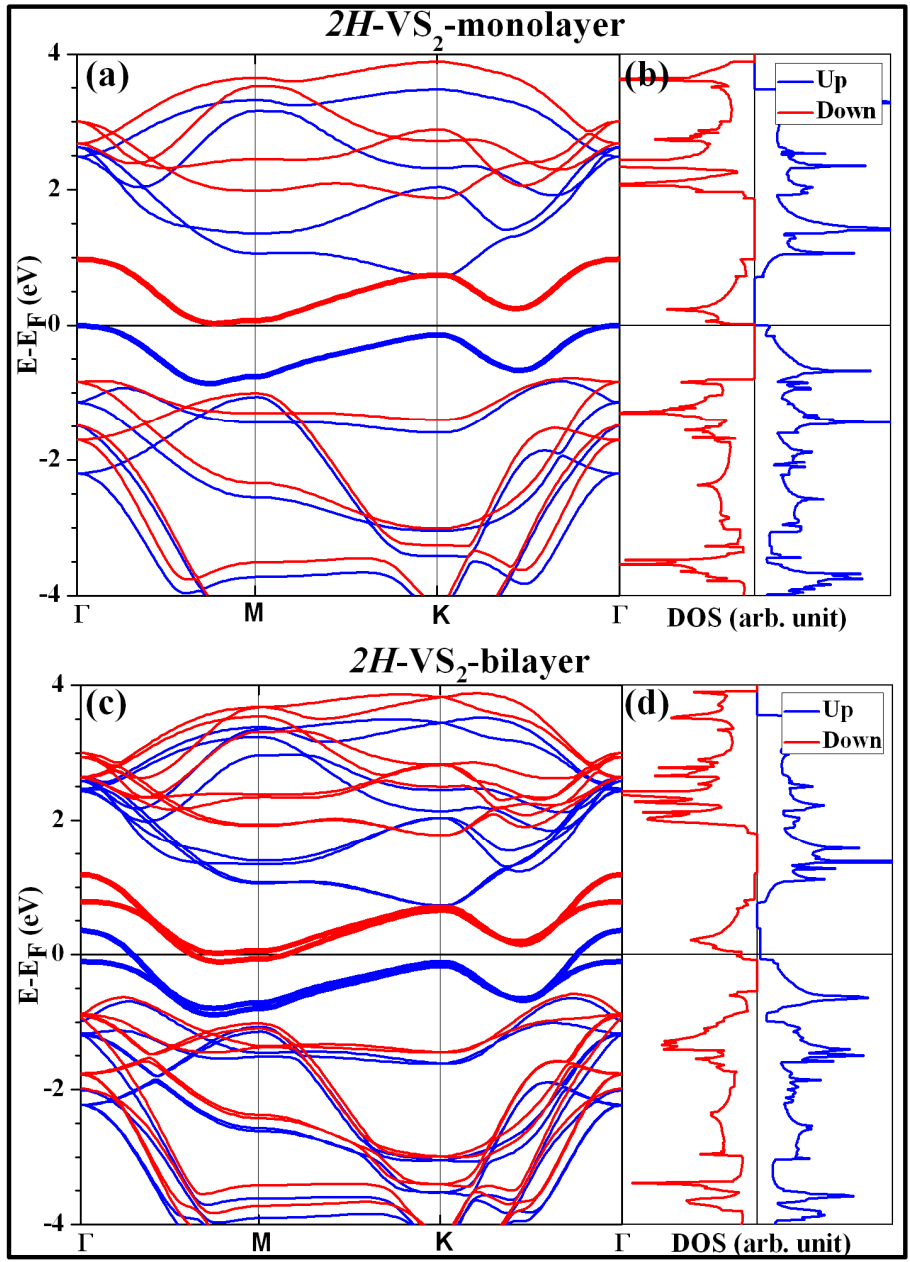

FIG. 3. Electronic band structure of *2H*-VS$_2$ (a) monolayer and (c) bilayer are represented in the figures along with their total DOS (b) and (d). Semimetal to metal transition occurred while going from monolayer to bilayer as clearly understood from the highlighted bands.

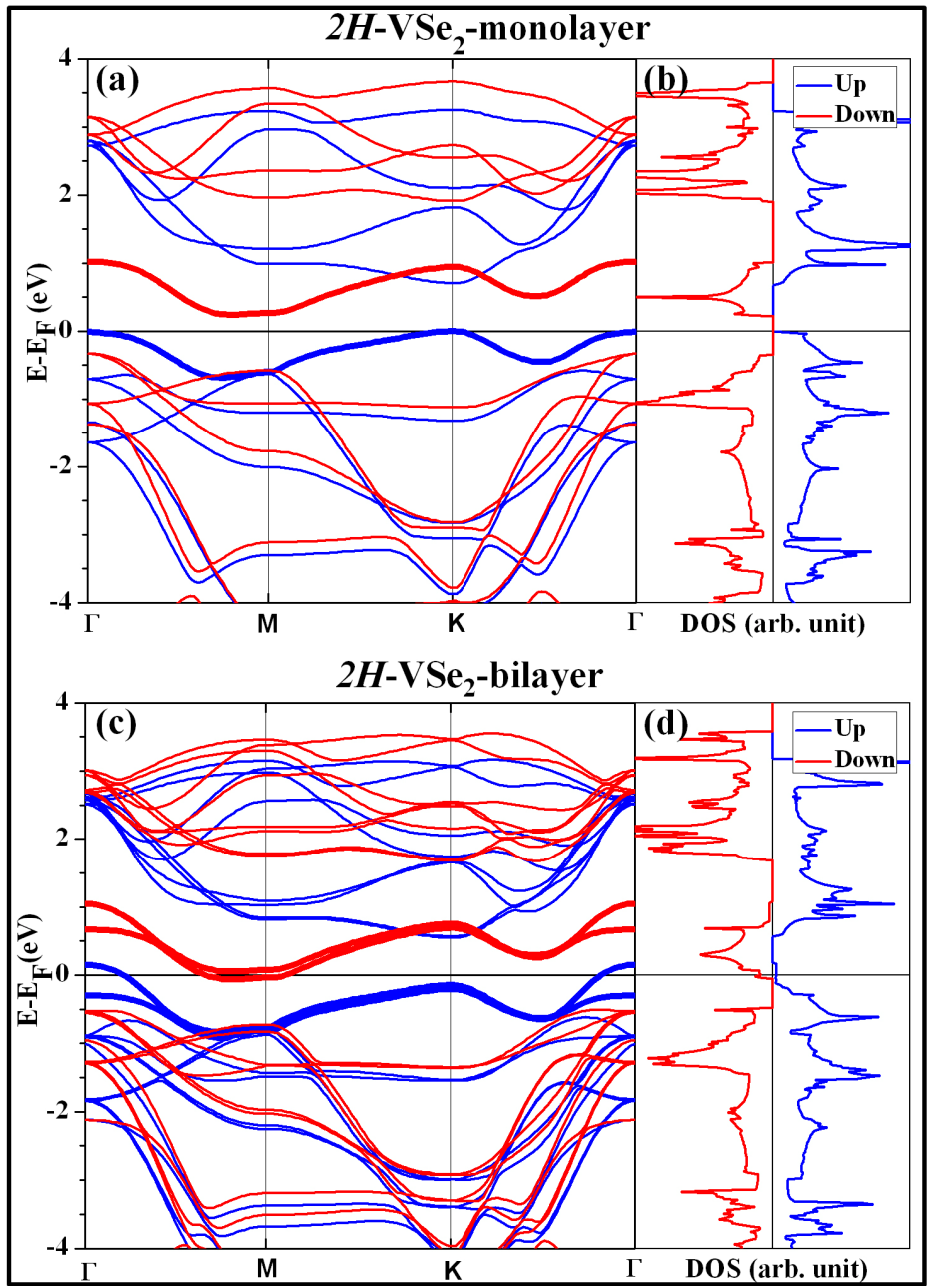

FIG. 4. Electronic band structure of *2H*-VSe$_2$ (a) monolayer and (c) bilayer are represented in the figures along with their total DOS (b) and (d). Semiconductor to metal transition occurred while going from monolayer to bilayer as clearly understood from the highlighted bands.

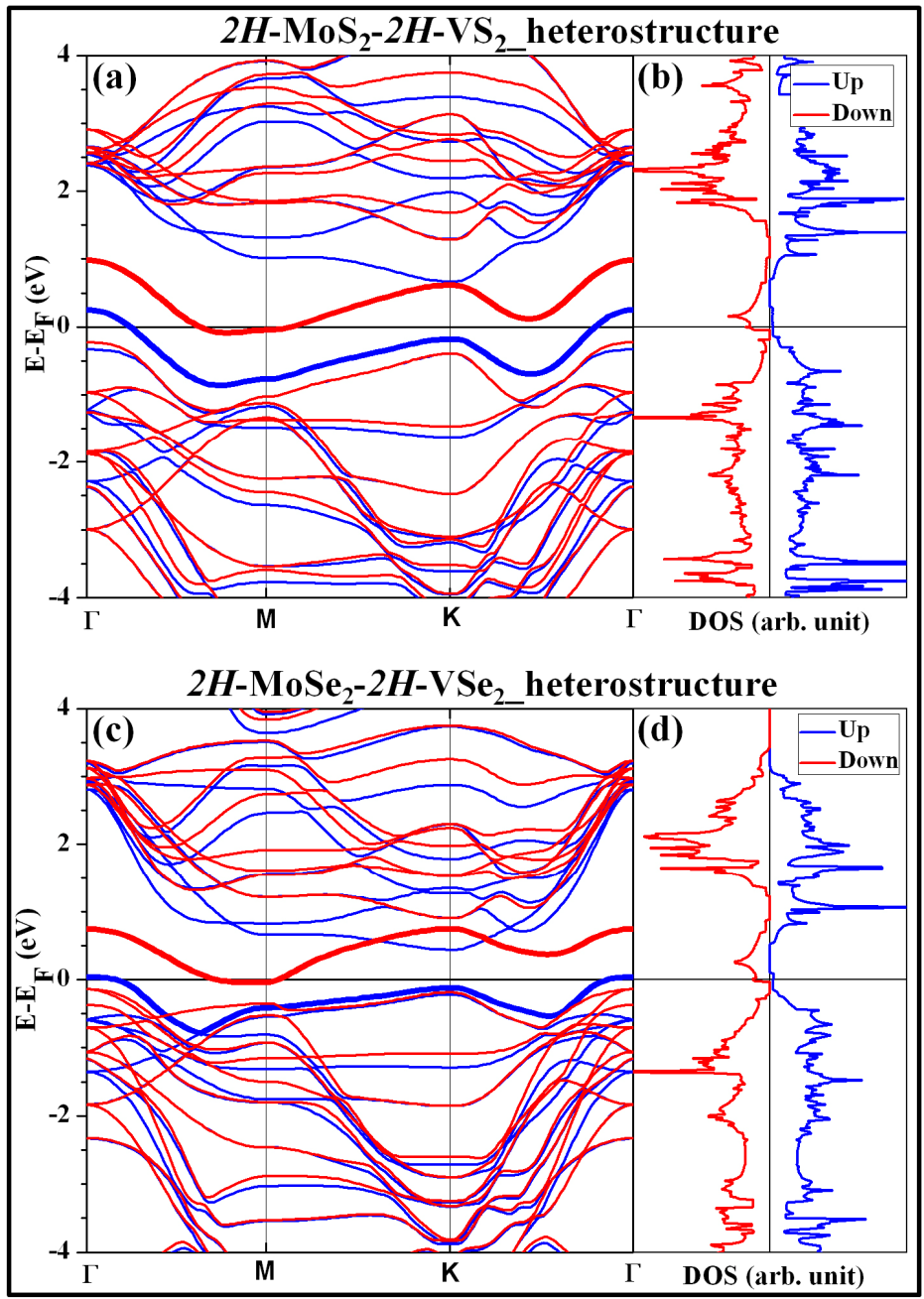

FIG. 5. Electronic band structure of (a) *2H*-MoS$_2$-*2H*-VS$_2$ (c) *2H*-MoSe$_2$-*2H*-VSe$_2$-heterobilayer are represented in the figures along with their total DOS (b) and (d). Here both the heterostructures show metallic electronic structure as revealed by their highlighted bands.

# APPENDIX

Self-consistent calculations have been carried out on the bulk phases of the $VX_2$ (X=S, Se, Te) systems discussed in this paper. The projected density of states (PDOS) of $VX_2$ (X=S, Se) systems and energy band dispersions of $VTe_2$ monolayer and bilayer have been plotted. These results have been appended here. In order to ensure continuity of the essential physics in this paper, we have included these figures and table in the APPENDIX.

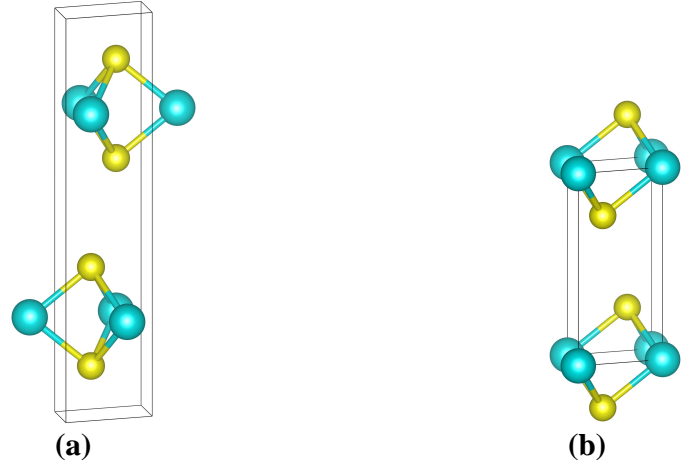

(a)  (b)

**FIG. A1:** Showing the $VX_2$-bulk crystal structure in (a) *2H*-phase and (b) *1T*-phase

**TABLE A1:** Estimated lattice parameters of *1T*-$VS_2$-bulk using different XC-functional with and without incorporating van der Waals (vdW) dispersion correction.

| XC-Funcional | *1T*-$VS_2$-bulk | |
|---|---|---|
| | $a$ (Å) | $c$ (Å) |
| LDA-CA | 3.10 | 5.54 |
| LDA-CA(+vdW) | 3.10 | 5.38 |
| GGA-PW91 | 3.18 | 6.50 |
| GGA-PW91(+vdW) | 3.18 | 5.89 |
| GGA-PBE | 3.18 | 6.60 |
| GGA-PBE(+vdW) | 3.17 | 5.91 |

Experimental* values: $a$=3.22 Å and $c$=5.76  [*G. A. Wiegers, *Physica* **19B**, 151 (1980) ]

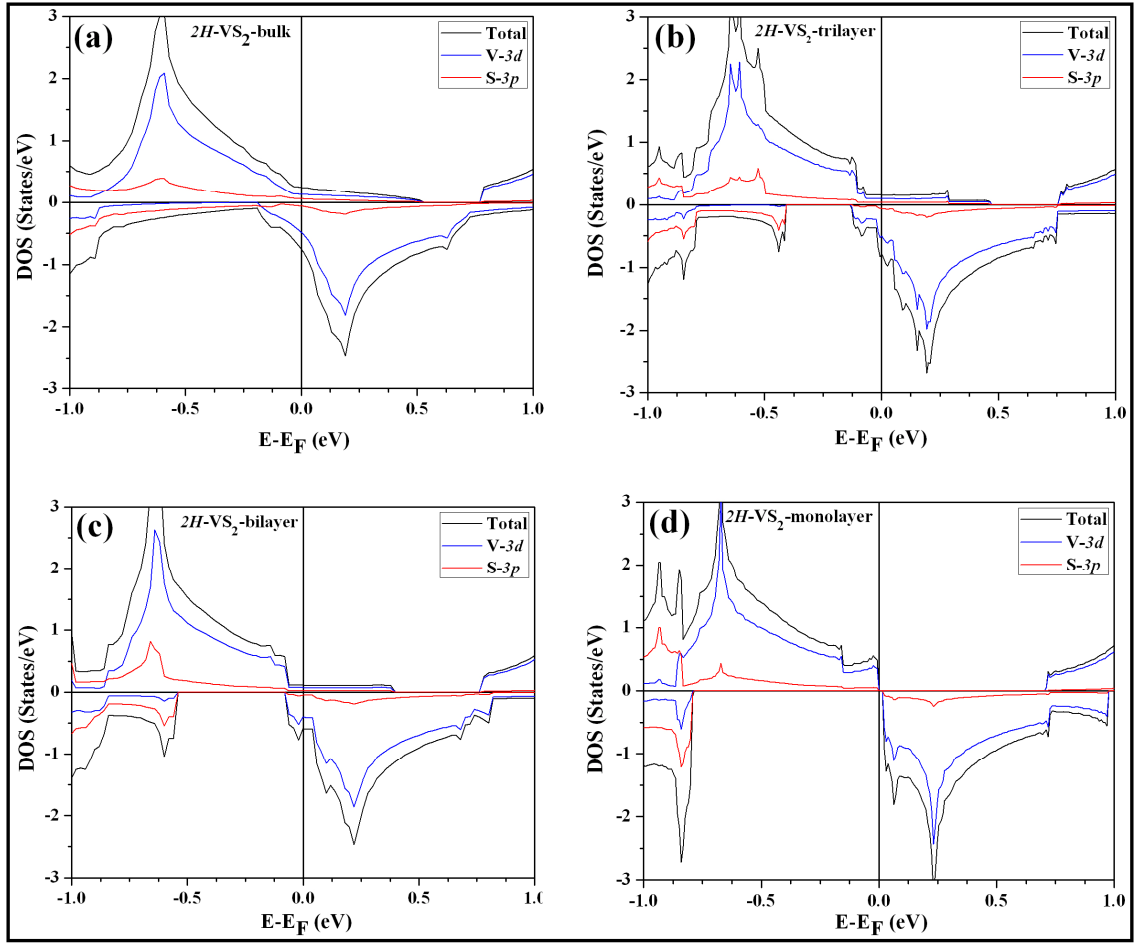

**FIG. A2:** Figure shows the site as well as orbital projected DOS of (a) bulk, (b) trilayer, (c) bilayer and (d) monolayer of *2H*-VS$_2$. This is a manifestation of metal to semimetal transition while down sizing the material from *2H*-VS$_2$-bulk to monolayer.

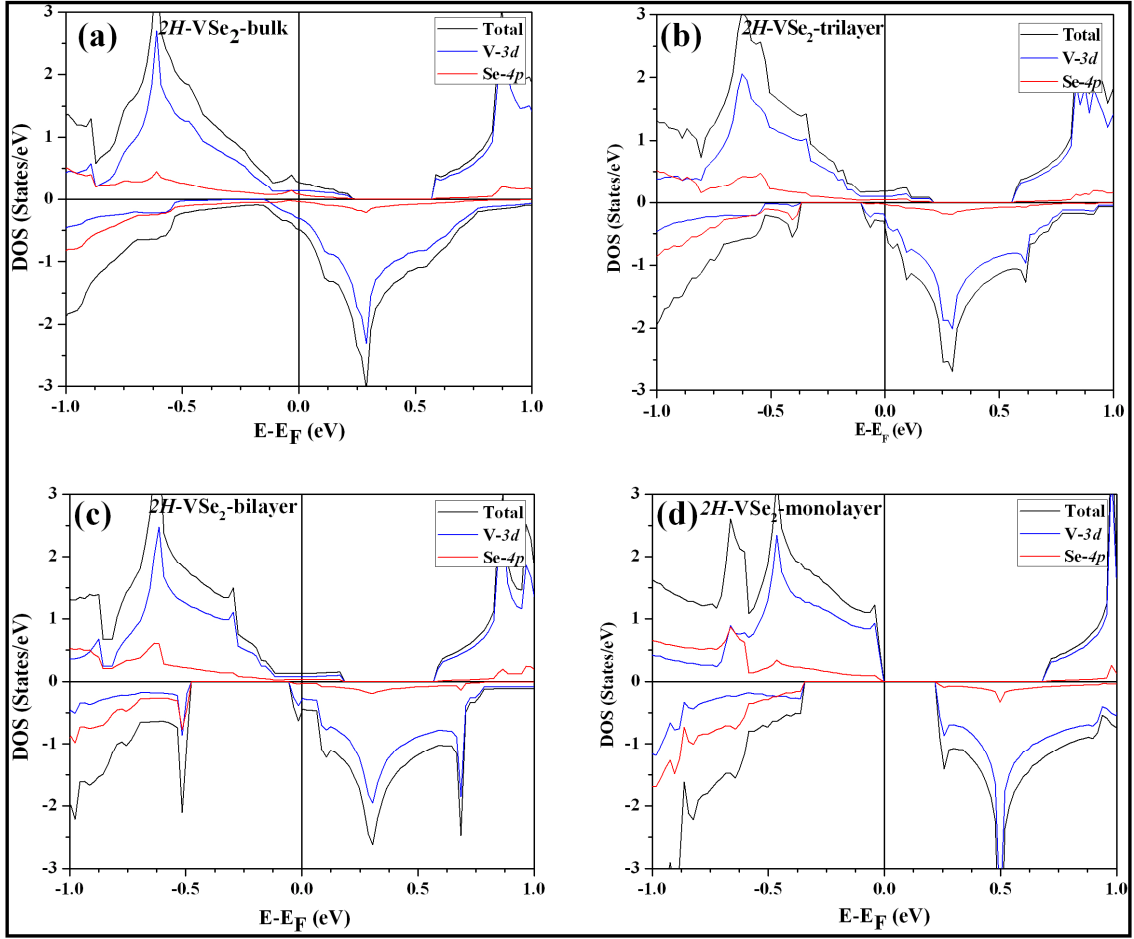

**FIG. A3:** Figure shows the site as well as orbital projected DOS of (a) bulk, (b) trilayer, (c) bilayer and (d) monolayer of *2H*-VSe$_2$. This is a manifestation of metal to narrow band gap semiconductor transition while downsizing the material from *2H*-VSe$_2$-bulk to monolayer.

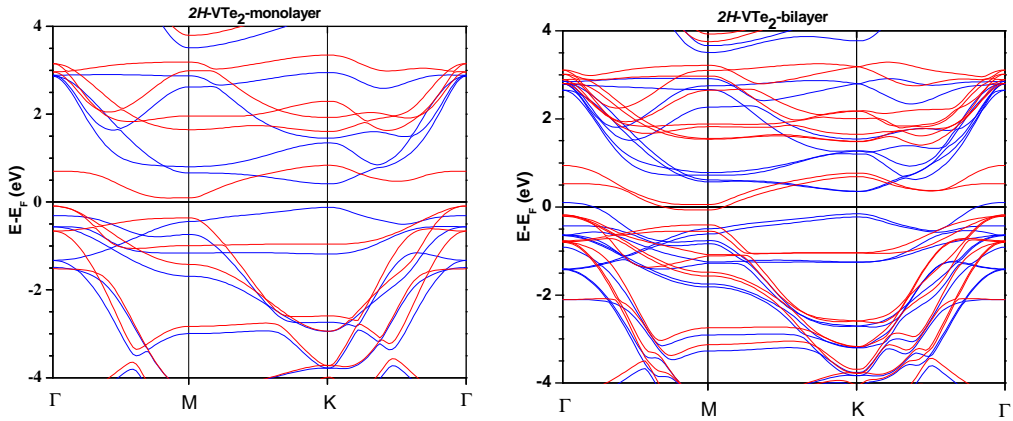

**FIG. A4:** Figures show energy band structure of *2H*-VTe$_2$ monolayer and bilayer.

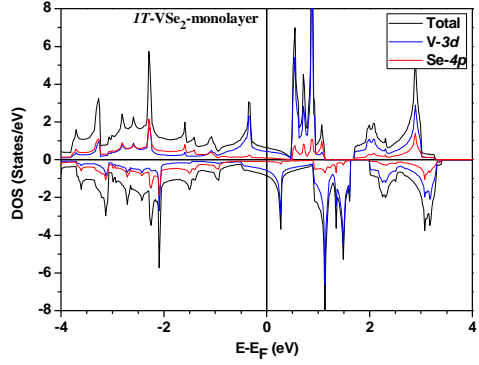

**FIG. A5:** Projected DOS of *1T*-VSe$_2$-monolayer

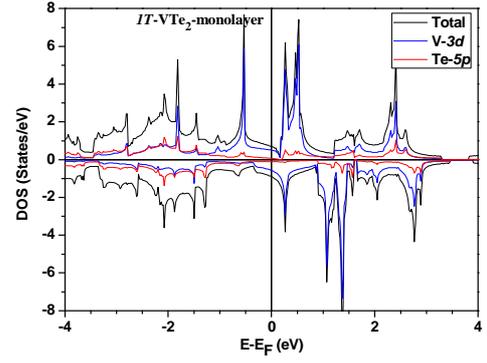

**FIG. A6**: Projected DOS of *1T*-VTe$_2$-monolayer

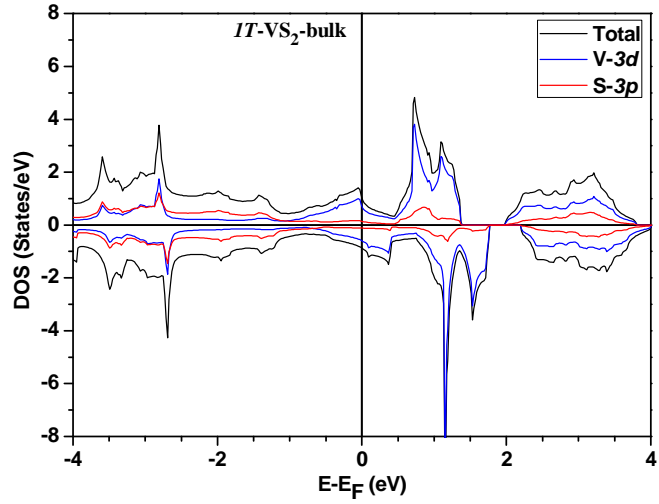

**FIG. A7:** Projected DOS of *1T*-VS$_2$-bulk

**Note on heterobilayers**: Estimated lattice parameters of *2H*-MoS$_2$ and *2H*-MoSe$_2$-monolayers are respectively 3.19 Å and 3.32 Å. These are having good lattice matching with *2H*-VS$_2$-monolayer (3.17 Å) and *2H*-VSe$_2$-monolayer (3.33 Å). Interface energies are found to be -0.1437 eV and -0.2085 eV respectively for *2H*-MoS$_2$-*2H*-VS$_2$ and *2H*-MoSe$_2$-*2H*-VSe$_2$ heterobilayers which are of the order of their corresponding *2H*-VX$_2$-bilayers.